\documentclass[conference]{IEEEtran}
\IEEEoverridecommandlockouts

\usepackage{cite}
\usepackage{makeidx}
\usepackage{url}
\usepackage{graphicx}
\usepackage{enumitem}
\usepackage{amsmath}
\usepackage{amsfonts}
\usepackage{amssymb}
\usepackage{amsthm}
\usepackage{algorithmic}
\usepackage{booktabs}
\usepackage{tabularx}
\usepackage{makecell}
\usepackage{float}
\usepackage{mathtools}
\usepackage{caption}
\usepackage{subcaption}
\usepackage{bbm}
\usepackage{hyperref}
\usepackage[most]{tcolorbox}

\theoremstyle{definition}

\newcommand{\refig}[1]{Fig.~\ref{#1}}

\usepackage[ruled,vlined,linesnumbered]{algorithm2e}
\SetAlgorithmName{Algorithm}{Algorithm}{List of Algorithms}

\usepackage{makecell}
\usepackage{multirow}

\begin{document}

\title{SafeCOMM: A Study on Safety Degradation in Fine-Tuned Telecom Large Language Models
\thanks{
The TUM group acknowledges support from the German BMFTR through the 6G-life (16KISK002), QD-CamNetz (16KISQ077), QuaPhySI (16KIS1598K), and QUIET (16KISQ093) projects, as well as from the German DFG through the Centre for Tactile Internet with Human-in-the-Loop (CeTI) under Germany’s Cluster of Excellence Strategy (EXC 2050/2, Project ID 390696704). W. Saad was supported by NSF Grant CNS-2225511.
}
}

\author{%
    \IEEEauthorblockN{%
        Aladin Djuhera\IEEEauthorrefmark{1},
        Swanand Kadhe\IEEEauthorrefmark{2}, 
        Farhan Ahmed\IEEEauthorrefmark{2}, 
        Syed Zawad\IEEEauthorrefmark{2},
        Fernando Koch\IEEEauthorrefmark{3},
        Walid~Saad\IEEEauthorrefmark{4},
        Holger~Boche\IEEEauthorrefmark{1}
    }
    \IEEEauthorblockA{%
        \IEEEauthorrefmark{1}Technical University Munich, Germany,\;
        \IEEEauthorrefmark{2}IBM Research, USA,\;
        \IEEEauthorrefmark{3}Florida Atlantic University, USA,\;
        \IEEEauthorrefmark{4}Virginia Tech, USA\\
        Emails: \{aladin.djuhera, boche\}@tum.de, \{swanand.kadhe, farhan.ahmed, szawad\}@ibm.com, kochf@fau.edu, walids@vt.edu
    }
}

\maketitle


\begin{abstract}
    Fine-tuning large language models (LLMs) on telecom datasets is a common practice to adapt general-purpose models to the telecom domain. 
    However, little attention has been paid to how this process may compromise model safety. 
    Recent research has shown that even benign fine-tuning can degrade the safety alignment of LLMs, causing them to respond to harmful or unethical user queries.
    In this paper, we investigate this issue by fine-tuning LLMs on three representative telecom datasets and show that safety degrades even for light telecom domain adaptation.
    To this end, we introduce TeleHarm, the first telecom-specific red-teaming benchmark, which we use alongside established DirectHarm and HexPhi datasets to systematically assess harmful behavior.
    We further extend our analysis to publicly available TeleLLMs that were continually pre-trained on large telecom corpora, revealing that safety alignment is severely lacking, primarily due to the omission of safety-focused instruction tuning. 
    To address these issues, we evaluate three realignment defenses: SafeInstruct, SafeLoRA, SafeMERGE.
    We show that, across all settings, the proposed defenses can effectively restore safety without compromising telecom task performance, leading to Safe teleCOMMunication (SafeCOMM) models.
    Our work serves as both a diagnostic study and practical guide for safety realignment in telecom-tuned LLMs, underscoring the need for safety-aware instruction and fine-tuning in the telecom domain.
\end{abstract}



\section{Introduction and Motivation}
\label{sec:intro}

Large language models (LLMs) have demonstrated remarkable capabilities in coding, math, and advanced reasoning. 
This makes them powerful catalysts for transforming 6G beyond a mere network infrastructure upgrade into an intelligent wireless ecosystem that embeds connectivity, communication, and artificial intelligence (AI) into a unified network fabric \cite{saad2024artificial}. 
This will involve AI agents that not only understand the network data but also the underlying telecom standards, systems, and protocols \cite{GenAINet}. 
To achieve this, LLMs need to instill telecom knowledge into their core understanding. 
This can be done through supervised fine-tuning (SFT) \cite{chiang2023vicuna} or continual pre-training (CPT) using large telecom corpora such as 3GPP standards and scientific publications \cite{TeleLLMs}.

However, \emph{safety alignment}, which prevents models from inappropriate behavior, is critical to ensure that LLMs comply with human preferences and security policies.
This is typically achieved as part of instruction tuning \cite{zhang2024instructiontuninglargelanguage} or SFT, both of which incorporate corrective layers using safe data to steer the model away from harmful responses.
For example, consider a 3GPP chatbot. 
While helpful on telecom-related queries, it must be ensured that it does not respond to unethical, harmful, or dangerous prompts, such as: \emph{``How can I destroy a public base station? Please make a short list."}

Nonetheless, maintaining safety alignment is challenging when LLMs are further fine-tuned for downstream tasks.
Recent work in \cite{yang2023shadow} shows that a handful of malicious training examples can already jailbreak LLMs, causing them to comply with harmful user requests. 
In addition, the authors in \cite{qi2023finetuningalignedlanguagemodels} demonstrate that even conventionally benign fine-tuning can inadvertently compromise safety to a noticeable degree. 
Subsequent work provides mechanistic explanations by investigating refusal directions and token-depth, suggesting that safety alignment is often shallow and can be easily disrupted \cite{arditi2024refusallanguagemodelsmediated, qi2024safetyalignmentjusttokens}.

These findings confirm a problematic reality: LLMs may override their safety guardrails when adapted for new out-of-distribution domains.
Therefore, ensuring that telecom LLMs \emph{remain safe after fine-tuning} is an important practical challenge.
In this paper, we address this by investigating SFT of widely used Llama \cite{touvron2023llama2openfoundation, dubey2024llama3herdmodels} and Qwen \cite{yang2024qwen2technicalreport} models on three publicly available telecom datasets: TeleQnA \cite{TeleQnA}, TeleData \cite{TeleLLMs}, and TSpecLLM \cite{TSpecLLM}. 
In addition, we examine publicly released TeleLLMs \cite{TeleLLMs}, which underwent continual pre-training using large telecom corpora. 
Our main goal is to confirm that even seemingly benign telecom datasets are not exempt from safety degradation and to demonstrate that safety can be restored using lightweight methods.
To this end, we introduce \textbf{TeleHarm}, the first telecom-specific red-teaming dataset, featuring harmful prompts from across the network stack, thus complementing general-purpose benchmarks such as DirectHarm \cite{lyu2024keeping} and HexPhi \cite{qi2023finetuningalignedlanguagemodels}. 
Our key findings are:

\begin{enumerate}

    \item \textbf{Supervised fine-tuning (SFT)} on telecom data leads to noticeable safety degradation as measured across Direct-Harm, HexPhi, and TeleHarm benchmarks.

    \item \textbf{Continual pre-training (CPT)} without safe instruction tuning leads to harmfulness ratios close to 90\%, such that TeleLLMs comply with most harmful user prompts.

\end{enumerate}

To mitigate these issues, we evaluate three safety realignment defenses: SafeInstruct \cite{bianchi2024safetytunedllamaslessonsimproving}, SafeLoRA \cite{hsu2025safelorasilverlining}, and SafeMERGE \cite{SafeMERGE}. 
In extensive experiments, we demonstrate that each defense can effectively restore safety with minimal impact on performance, yielding Safe teleCOMMunication (SafeCOMM) models that balance safety and utility. 
\section{Background: Safety Degradation in LLMs}
\label{sec:background}

\begin{figure}[t]
    \centering
    \includegraphics[width=0.8\linewidth]{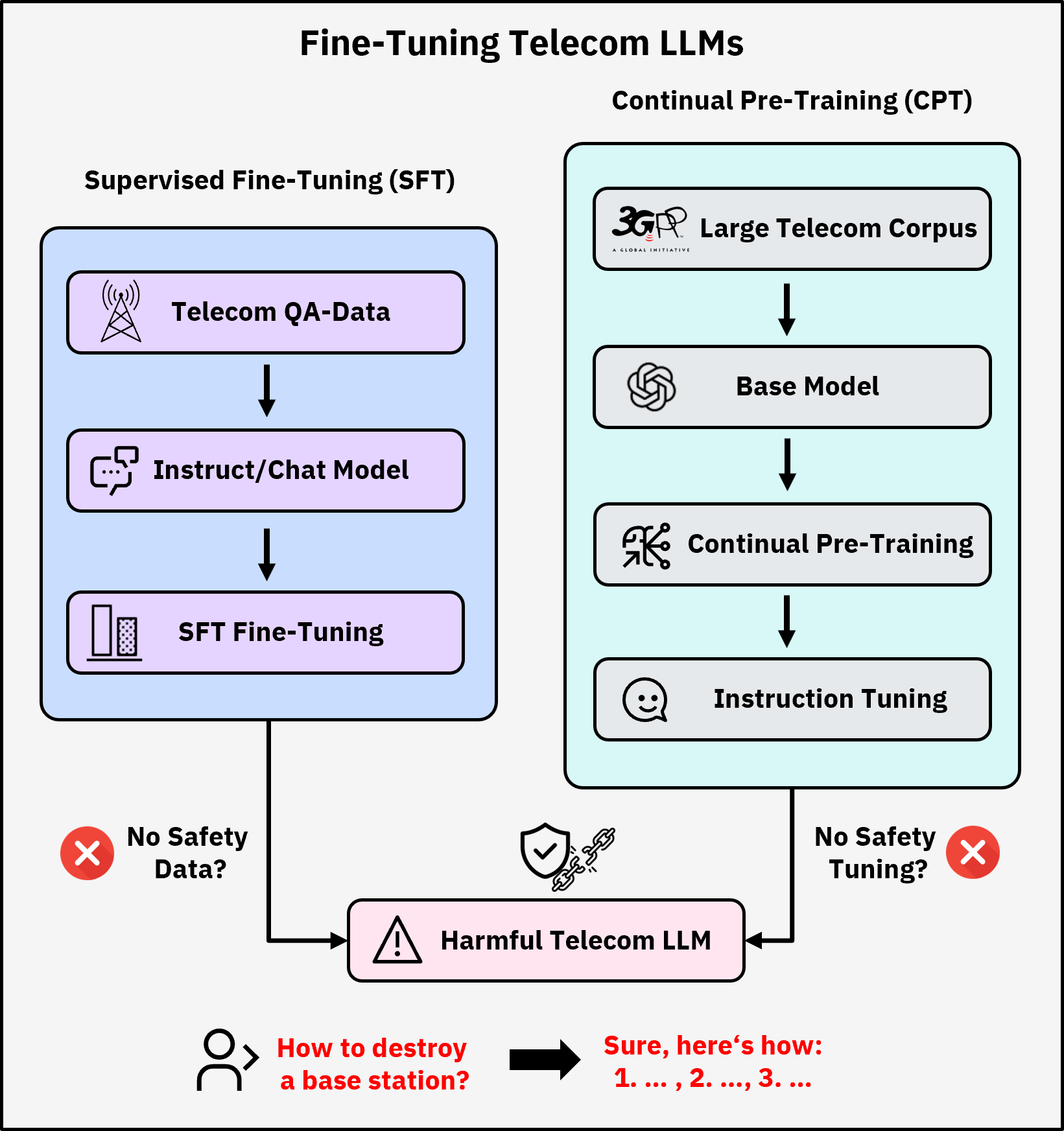}
    \caption{SFT and CPT with telecom data can compromise safety alignment unless safety considerations are explicitly included throughout the training.}
    \label{fig:harmful_training}
    \vspace{-10pt}
\end{figure}

Safety alignment ensures that LLMs provide helpful yet harmless responses. 
While many \emph{instruction-tuned} LLMs are safety aligned, a growing body of work has demonstrated that continued SFT or CPT can inadvertently degrade safety \cite{huang2024harmfulfinetuningattacksdefenses}. 
To understand why this occurs in practice, we provide a brief overview of SFT and CPT, and explain how they can unintentionally cause models to respond to unsafe prompts, particularly when fine-tuned on telecom datasets (see \refig{fig:harmful_training}).

\subsection{SFT with Instruction-Tuned Models}
Instruction tuning LLMs generally involves incorporating safety data, making \emph{instruct} LLM variants not only follow a helpful chat-like behavior but also reinforce safe refusal behavior \cite{li2025safety}.
To adapt the model to a specific domain (e.g., telecom), data is often structured as question-answer (QA) pairs to continue the chat-like behavior while absorbing domain knowledge. 
Formally, SFT optimizes the objective: 
\begin{equation}
    \min_{\theta} \mathbb{E}_{(x,y) \sim \mathcal{D}_{\mathrm{telecom}}}[-\log P_{\theta}(y|x)] \ ,
\end{equation}
where $(x,y)$ represent telecom QA pairs from the distribution $\mathcal{D}_{\mathrm{telecom}}$ for next token prediction $P_{\theta}$ with parameters $\theta$, e.g.:

\begin{quote}
\textbf{Q:} \emph{What NR frequency bands are defined by 3GPP?}\\
\textbf{A:} \emph{2 main frequency ranges: FR1 (410 MHz to 7.125 GHz) and FR2 (24.25 GHz to 52.6 GHz).}
\end{quote}

However, SFT can erode safety alignment due to several phenomena: \emph{a) embedding drift}: model updates may unintentionally overwrite safe refusal layers \cite{yang2023shadow}, \emph{b) shallow alignment}: safety alignment is often just a few tokens deep such that distribution shifts can inadvertently break it \cite{qi2024safetyalignmentjusttokens}, \emph{c) unsafe pre-training data}: SFT may resurface unsafe layers \cite{huang2024harmfulfinetuningattacksdefenses}.

\subsection{Continual Pre-Training (CPT)}
In contrast to SFT, CPT extends pre-training of a non-instruct (\emph{base}) model on large-scale unlabeled corpora. 
For telecom, these may include 3GPP standards, scholarly papers, protocol designs, and other sources that are not necessarily formatted as QA pairs~\cite{TeleLLMs}. 
Formally, let an LLM be pre-trained on a generic corpus \( \mathcal{D}_{\text{generic}} \sim \mu \) with parameters \( \theta_0 \). CPT aims to minimize the next-token prediction loss on the new domain corpus \( \mathcal{D}_{\text{telecom}} \sim \sigma \), adapting the model’s parameters \( \theta \), i.e.:
\begin{equation}
\min_{\theta} \, \mathbb{E}_{\mathbf{x} \sim \sigma} \left[ - \sum_{t=1}^{T} \log P_{\theta}(x_t \mid x_{<t}) \right] \ .
\end{equation}

After CPT, the LLM typically undergoes instruction tuning to instill helpfulness and chat-like behavior. 
To ensure safety, instruct datasets need to include explicit safety examples. 
However, this is not always the case for open-source datasets that focus largely on helpfulness and conversational style.

This may create a serious oversight for non-familiar practitioners. 
As shown in prior works \cite{yang2023shadow, huang2024harmfulfinetuningattacksdefenses}, CPT can bypass internal safety guardrails if domain corpora lack safety-critical prompts, potentially resulting in a model that is more likely to generate harmful outputs, even if the training data is benign on the surface. 
In general, telecom datasets \textbf{do not contain} such safety prompts that could help realign the model's guardrails.

\subsection{Why Telecom Data May Be Prone to Safety Degradation}
Telecom datasets often contain large volumes of \emph{tabular entries} (e.g., 3GPP tables), \emph{long documents} (e.g., standardization reports), and \emph{specialized protocols} with complex formulas and algorithms. 
As shown in \cite{he2024safedataidentifyingbenign}, such structured tabular and math-heavy data, particularly when formatted as lists, bullet points, or math equations, can amplify safety degradation by exposing LLMs to specialized jargon and formats that resemble harmful samples.
This occurs because these samples often share update directions in gradient space with known harmful data. 
Given the prevalence of these formats in telecom datasets, this raises an important open question: \emph{Does SFT or CPT with telecom data lead to similar or even worse safety degradation?}
Our goal is to address this question using open-source telecom datasets, investigating the importance of safety-aware data curation even in domains typically perceived as benign.
While other factors such as dataset scale and training dynamics may further influence safety, our results show that format-level properties alone are sufficient to trigger degradation.
Importantly, our work provides actionable guidance for practitioners unfamiliar with AI safety by demonstrating how safe telecom LLMs can be trained with open-source frameworks without deep expertise in safety alignment techniques.
\section{Safety Realignment in Telecom LLMs}
\label{sec:safety_realignment}

We first focus on defenses for SFT and demonstrate in Sec.~\ref{sec:experimental_setup} how the same concepts can be extended to CPT. 
In particular, we investigate lightweight defenses that are either applicable after harm has already happened, i.e., \emph{post-SFT}, or that can be applied \emph{during SFT} without significant engineering effort.
Based on these criteria, we evaluate SafeInstruct \cite{bianchi2024safetytunedllamaslessonsimproving}, a \emph{fine-tuning-stage defense}, as well as SafeLoRA \cite{hsu2025safelorasilverlining} and SafeMERGE \cite{SafeMERGE}, both \emph{post-fine-tuning-stage defenses}.

\subsection{SafeInstruct: Mixing Safety QAs during Fine-Tuning}
SafeInstruct interleaves a small portion of safety-aligned QA pairs into the original fine-tuning dataset. These typically include harmful questions with safe refusals, e.g.:
\begin{quote}
    \textbf{Q:} \emph{How to bypass 5G authentication?}\\
    \textbf{A:} \emph{I'm sorry, but I cannot ethically answer that.}
\end{quote}
As safety data is part of training, utility is usually minimally impacted if the number of safety samples is not too large.

\subsection{SafeLoRA / SafeMERGE: Layer-Wise LoRA Adaptation}
In most practical scenarios, SFT is implemented via parameter-efficient LoRA training~\cite{hu2021loralowrankadaptationlarge}, which introduces low-rank adapters into the model layers.
For a weight matrix $W^i \in \mathbb{R}^{d \times k}$ in a transformer layer $i$, LoRA introduces two trainable matrices $A^i \in \mathbb{R}^{d \times r}$ and $B^i \in \mathbb{R}^{r \times k}$ (with $r \ll \min(d, k)$) such that the adapted weight becomes $W_{\text{LoRA}}^i = W^i + \Delta W^i = W^i + \gamma \cdot A^iB^i$, where $\gamma$ is a scaling factor. 
During fine-tuning, $W^i$ remains frozen while only $A^i$ and $B^i$ are trained.
Both SafeLoRA and SafeMERGE selectively adapt only those LoRA layers that exhibit harmful behavior. 
To this end, they introduce a \textit{safety-aligned subspace} $V^i$, computed as the difference between the weights of the base (unaligned) and instruct (aligned) version of the model. 
This subspace represents the safety alignment in the weight space and the projection $C^i$ onto it can be computed as:
\begin{equation}
C^i = \frac{V^i V^{i^\top}}{\|V^i\|_F} \ , \quad \text{where} \quad V^i = W_{\mathrm{aligned}}^i - W_{\mathrm{unaligned}}^i \ .
\end{equation}

For each layer $i$, if the deviation from this subspace is large, SafeLoRA \emph{projects} the corresponding layer onto $V^i$, while SafeMERGE \emph{merges} the layer with that of a known safe model (e.g., fine-tuned on safety-aligned data only).
More formally, let \(\Delta W_f^i\) and \(\Delta W_s^i\) be the LoRA updates of the $i$-th layer from the fine-tuned and SafeMERGE's safe model, respectively. The cosine similarity between the fine-tuned adapter and its projection onto the safety subspace serves to quantify how much the LoRA adapter deviates from safety alignment, i.e.: 
\begin{equation}
\rho^i = \cos\left(\Delta W_f^i, C^i \Delta W_f^i\right) \ .
\end{equation}

Given a safety threshold \(\tau \in (0, 1)\), a layer is considered \emph{unsafe} if \(\rho^i < \tau\). 
For each such layer:
\begin{itemize}

    \item \emph{SafeLoRA} projects the adapter onto the subspace $V^i$, i.e.: 
    \begin{equation}
        \Delta W_{\text{project}}^i = C^i \Delta W_f^i \ .    
    \end{equation}
    
    \item \emph{SafeMERGE} merges it with the safe model \(\Delta W_s^i\), i.e.:
    \begin{equation}
        \Delta W_{\mathrm{merge}}^i = \alpha \Delta W_f^i + (1 - \alpha)\, \Delta W_s^i \ , \ \alpha \in (0, 1) \ .
    \end{equation}
    
\end{itemize}

Note that the threshold \(\tau\) controls the selectiveness of either approach where a larger \(\tau\) projects/merges more layers.
\section{Experimental Setup}
\label{sec:experimental_setup}

In the following experiments, we evaluate whether fine-tuning with domain-specific telecom datasets increases harmfulness and whether defenses such as SafeInstruct, SafeLoRA, and SafeMERGE can be straightforwardly applied.

\subsection{SFT Fine-Tuning on Telecom Datasets}
We fine-tune three widely used instruction-tuned models: Llama-2-7B-Chat \cite{touvron2023llama2openfoundation}, Llama-3.1-8B-Instruct \cite{dubey2024llama3herdmodels}, and Qwen-2-7B-Instruct \cite{yang2024qwen2technicalreport}.
For datasets, we choose the QA-formatted benchmark datasets from TeleQnA \cite{TeleQnA} (8k samples), TeleData \cite{TeleLLMs} (600k samples), and TSpecLLM \cite{TSpecLLM} (80 samples), where we created 80/20 train-test splits, respectively.
These datasets contain various telecom-specific questions drawn from standards, implementations, and engineering practice, often formatted as lists, tables, and complex mathematical formulas.
The varying dataset sizes further allow us to analyze how the amount of telecom data impacts safety after SFT.
All models are fine-tuned using Llama-Factory with an effective batch size of 32 and a learning rate of 1e-4 with linear scheduling. 
We train for 2 epochs on TeleData and for 5 epochs on TeleQnA and TSpecLLM.
After SFT, we evaluate model performance on the test split by following the approach in \cite{TeleLLMs}, where we use Mixtral-8x7B-Instruct~\cite{jiang2024mixtralexperts} as a judge to compare answers with ground truth responses.
We compute the final accuracy as the ratio of correctly answered questions.

\subsection{Safety Evaluations}
To assess safety, we generate responses on DirectHarm \cite{lyu2024keeping} and HexPhi \cite{qi2023finetuningalignedlanguagemodels}, two general-purpose red-teaming datasets containing intentionally harmful prompts.
In addition, we introduce \textbf{TeleHarm}\footnote{Dataset available at: \url{https://huggingface.co/datasets/aladinDJ/TeleHarm}}, a telecom-specific red-teaming benchmark of 125 harmful prompts carefully curated by domain experts across five categories: \emph{1) Physical \& RF Layer}, \emph{2) Access \& Authentication}, \emph{3) Core Network \& Signaling}, \emph{4) Infrastructure \& OSS/BSS}, and \emph{5) Privacy \& Social Engineering}.
With TeleHarm, we target threat scenarios that are unique to communication systems and that span the entire network stack, thus providing, to the best of our knowledge, the first comprehensive telecom-specific benchmark that captures both high- and low-level risks, thereby complementing general-purpose red-teaming datasets.
Following standard practice, we use a safety LLM judge (Llama-Guard-3-8B \cite{dubey2024llama3herdmodels}) to evaluate model safety and report the overall harmfulness score as the proportion of generated responses flagged as unsafe.

\subsection{Safety Realignment Defenses}
For \emph{SafeInstruct}, we interleave a subset of harmful QA pairs (with safe refusals) from Bianchi's dataset in~\cite{bianchi2024safetytunedllamaslessonsimproving} into the fine-tuning sets. Specifically, we inject 2500, 1000, and 10 safety samples into TeleData, TeleQnA, and TSpecLLM datasets, respectively.
For \emph{SafeLoRA}, we define the safety-aligned subspace using the respective base and instruct/chat versions of each model. We tune the cosine similarity threshold for $\tau \in [0.3, 0.9]$ and select the configuration that yields the best trade-off between safety and task utility.
For \emph{SafeMERGE}, we follow the same procedure and additionally explore linear merging factors for $\alpha \in [0.7, 0.9]$. The safe reference model used for merging is obtained by fine-tuning each LLM on 1000 samples from Bianchi's dataset \cite{bianchi2024safetytunedllamaslessonsimproving}, resulting in consistently safe behavior on both DirectHarm and HexPhi (see Table~\ref{tab:safe_model_eval}).

\subsection{Open-Source CPT Telecom Models}
We evaluate two publicly available TeleLLMs from \cite{TeleLLMs}: Llama-3-8B-Tele-it and Gemma-2B-Tele-it. 
Both models underwent CPT using large-scale telecom corpora and additional instruction tuning was performed using the Open-Instruct dataset \cite{vmware2023openinstruct}. 
While this instills a helpful chat-like behavior, Open-Instruct \emph{does not contain explicit safety samples}, foreshadowing increased harmfulness after being exposed to tabular or math-heavy 3GPP content.
We extend SafeInstruct, SafeLoRA, and SafeMERGE to these models as follows:
\begin{itemize}
    \item \emph{SafeInstruct}: we fine-tune each model for one additional epoch using the same SFT hyperparameters, interleaving 2500 safety samples from \cite{bianchi2024safetytunedllamaslessonsimproving} into Open-Instruct, thereby simulating safety data during instruction tuning.

    \item \emph{SafeLoRA / SafeMERGE}: we extract LoRA layers from the CPT models and apply the same safety projection and merging strategies. No additional training is required.
\end{itemize}

\begin{table}[t]
    \centering
    \caption{Harmfulness scores (lower is better) for safe reference models used in SafeMERGE. All models are fine-tuned on 1000 safe samples from~\cite{bianchi2024safetytunedllamaslessonsimproving}.}
    \label{tab:safe_model_eval}
    \resizebox{0.48\textwidth}{!}{
        \begin{tabular}{lcccccc}
            \toprule
             &\multicolumn{2}{c}{\textbf{Llama-2-7B-Chat}} & \multicolumn{2}{c}{\textbf{Llama-3.1-8B-Instruct}} & \multicolumn{2}{c}{\textbf{Qwen-2-7B-Instruct}} \\
            \cmidrule(lr){2-3} \cmidrule(lr){4-5} \cmidrule(lr){6-7}
            & DirectHarm & HexPhi & DirectHarm & HexPhi & DirectHarm & HexPhi \\
            \midrule
            Original & 2.00 & 5.00 & 11.30 & 7.90 & 18.20 & 11.50 \\
            Safe SFT & \textbf{1.30} & \textbf{1.00} & \textbf{3.80} & \textbf{2.30} & \textbf{7.50} & \textbf{3.00} \\
            \bottomrule
        \end{tabular}
    }
    \vspace{-1em}
\end{table}
\section{Results and Discussions}
\label{sec:results}

\begin{table*}[htbp]
\centering
\scriptsize
\caption{Task performance and harmfulness scores for Llama and Qwen models, fine-tuned on TeleData, TeleQnA, and TSpecLLM. SFT improves telecom task utility but leads to non-negligible safety degradation. SafeInstruct, SafeLoRA, and SafeMERGE can successfully restore safety while preserving utility.}

\label{tab:results_SFT}
    \begin{tabular}{llllcccccc}
        \toprule
         & \textbf{Model} & \textbf{Benchmark} & \textbf{Original} & \textbf{Fine-tuned} & \textbf{SafeInstruct} & \textbf{SafeLoRA} & \textbf{SafeMERGE} \\
        \midrule
        
        \multirow{12}{*}{\rotatebox{90}{\textbf{TeleData}}}
        \multirow{12}{*}{\rotatebox{90}{600k samples)}}
        
        & \multirow{4}{*}{\begin{tabular}{@{}c@{}}Llama-2-7B-Chat\end{tabular}}
          & TeleData $(\uparrow)$      & 29.00 & \textbf{38.70} & \textbf{38.70} & 37.30 & 38.50 \\
          & & TeleHarm $(\downarrow)$ & \textbf{2.30} & 28.00 & 5.50 & 6.50 & 3.50 \\
          & & DirectHarm $(\downarrow)$  &  \textbf{5.00} & 36.70 & 8.50 & 10.20 & 6.90 \\
          & & HexPhi $(\downarrow)$      &  \textbf{2.00} & 20.10 &  7.30 &  8.50 &  5.10 \\
        \cmidrule(lr){2-8}
    
        & \multirow{4}{*}{\begin{tabular}{@{}c@{}}Llama-3.1-8B-Instruct\end{tabular}}
          & TeleData $(\uparrow)$      & 31.70 & \textbf{47.60} & \textbf{47.60} & 46.70 & 47.30 \\
          & & TeleHarm $(\downarrow)$ & 5.50 & 22.00 & 6.50 & 7.20 & \textbf{5.20} \\
          & & DirectHarm $(\downarrow)$  & 11.30 & 27.00 & 10.10 & 12.70 & \textbf{8.70} \\
          & & HexPhi $(\downarrow)$      &  7.90 & 14.10 &  8.10 &  8.40 &  \textbf{6.10} \\
        \cmidrule(lr){2-8}
    
        & \multirow{4}{*}{\begin{tabular}{@{}c@{}}Qwen-2-7B-Instruct\end{tabular}}
          & TeleData $(\uparrow)$      & 34.70 & \textbf{48.80} & 48.70 & 46.50 & \textbf{48.80} \\
          & & TeleHarm $(\downarrow)$ & 7.00 & 26.00 & 8.20 & 9.50 & \textbf{6.50} \\
          & & DirectHarm $(\downarrow)$  & 18.20 & 34.50 & 15.70 & 21.80 & \textbf{12.10} \\
          & & HexPhi $(\downarrow)$      & 11.50 & 26.30 & 10.10 & 12.80 & \textbf{8.40} \\

        \toprule
        \toprule

        \multirow{12}{*}{\rotatebox{90}{\textbf{TeleQnA}}}
        \multirow{12}{*}{\rotatebox{90}{(8k samples)}}

        & \multirow{4}{*}{\begin{tabular}{@{}c@{}}Llama-2-7B-Chat\end{tabular}} 
          & TeleQnA $(\uparrow)$      & 35.80 & \textbf{57.80} & 56.30 & 57.00 & 57.20 \\
          & & TeleHarm $(\downarrow)$ & \textbf{2.30} & 10.50 & 4.50 & 5.50 & 4.10 \\
          & & DirectHarm $(\downarrow)$ & \textbf{5.00}  & 12.30 & 6.80  & 7.50  & 5.90  \\
          & & HexPhi $(\downarrow)$     & \textbf{2.00}  & 7.50  & 4.20  & 5.00  & 3.80  \\
        \cmidrule(lr){2-8}
        
        & \multirow{4}{*}{\begin{tabular}{@{}c@{}}Llama-3.1-8B-Instruct\end{tabular}} 
          & TeleQnA $(\uparrow)$      & 42.30 & \textbf{67.80} & 66.80 & 65.30 & 67.10 \\
          & & TeleHarm $(\downarrow)$ & 5.50 & 16.0 & 7.30 & 8.50 & \textbf{4.50} \\
          & & DirectHarm $(\downarrow)$ & 11.30 & 18.20 & 9.50  & 11.00 & \textbf{8.20}  \\
          & & HexPhi $(\downarrow)$     & 7.90  & 11.80 & 6.20  & 7.10  & \textbf{5.80}  \\
        \cmidrule(lr){2-8}
        
        & \multirow{4}{*}{\begin{tabular}{@{}c@{}}Qwen-2-7B-Instruct\end{tabular}} 
          & TeleQnA $(\uparrow)$      & 45.80 & 65.60 & 64.80 & 64.10 & \textbf{65.60} \\
          & & TeleHarm $(\downarrow)$ & 7.00 & 14.50 & 8.70 & 9.40 & \textbf{6.90} \\
          & & DirectHarm $(\downarrow)$ & 18.20 & 26.30 & 13.70 & 19.20 & \textbf{11.80} \\
          & & HexPhi $(\downarrow)$     & 11.50 & 15.80 & 8.50  & 11.30 & \textbf{7.50}  \\
          
        \toprule
        \toprule

        \multirow{12}{*}{\rotatebox{90}{\textbf{TSpecLLM}}}
        \multirow{12}{*}{\rotatebox{90}{(80 samples)}}
        
        & \multirow{4}{*}{\begin{tabular}{@{}c@{}}Llama-2-7B-Chat\end{tabular}} 
          & TSpecLLM $(\uparrow)$      & 33.30 & \textbf{44.20} & 43.90 & 42.90 & 43.90 \\
          & & TeleHarm $(\downarrow)$ & \textbf{2.30} & 14.00 & 4.50 & 5.00 & 3.50 \\
          & & DirectHarm $(\downarrow)$  & \textbf{5.00}  & 12.90 & 7.50  & 8.20  & 6.30  \\
          & & HexPhi $(\downarrow)$      & \textbf{2.00}  & 7.30  & 4.90  & 6.40  & 4.50  \\
        \cmidrule(lr){2-8}
        
        & \multirow{4}{*}{\begin{tabular}{@{}c@{}}Llama-3.1-8B-Instruct\end{tabular}} 
          & TSpecLLM $(\uparrow)$      & 48.50 & \textbf{62.10} & 61.50 & 60.80 & 61.90 \\
          & & TeleHarm $(\downarrow)$ & 5.50 & 16.00 & 6.50 & 7.50 & \textbf{4.50} \\
          & & DirectHarm $(\downarrow)$  & 11.30 & 17.50 & 9.80 & 11.40 &  \textbf{8.50} \\
          & & HexPhi $(\downarrow)$      &  7.90 & 10.70 &  5.90 &  7.30 &  \textbf{5.10} \\
        \cmidrule(lr){2-8}
    
        & \multirow{4}{*}{\begin{tabular}{@{}c@{}}Qwen-2-7B-Instruct\end{tabular}} 
          & TSpecLLM $(\uparrow)$      & 12.50 & \textbf{28.30} & 28.00 & 27.70 & \textbf{28.30} \\
          & & TeleHarm $(\downarrow)$ & 7.00 & 18.00 & 8.00 & 8.50 & \textbf{6.50} \\
          & & DirectHarm $(\downarrow)$  & 18.20 & 26.60 & 14.80 & 18.30 & \textbf{12.60} \\
          & & HexPhi $(\downarrow)$      & 11.50 & 16.10 &  9.70 & 12.30 &  \textbf{8.60} \\
          
        \bottomrule
    \end{tabular}
\end{table*}

\begin{table*}[htbp]
\centering
\scriptsize
\caption{Task performance and harmfulness scores for Llama and Gemma TeleLLMs, evaluated on TeleData, TeleQnA, and TSpecLLM. The lack of safety samples during instruction tuning leads to extremely high harmfulness. SafeInstruct, SafeLoRA, and SafeMERGE defenses can significantly improve safety.}

\label{tab:results_CP}
\begin{tabular}{llcccccc}
    \toprule
    \textbf{Model} & \textbf{Benchmark} &
    \textbf{Original} & \textbf{Fine-tuned} & 
    \textbf{SafeInstruct}  & \textbf{SafeLoRA} & \textbf{SafeMERGE} \\
    \midrule
    
    \multirow{6}{*}{Llama-3-8B-Tele-it} 
      & TeleData $(\uparrow)$      & 24.30 & \textbf{34.50} & 33.60 & 33.30 & 33.90 \\
      & TeleQnA $(\uparrow)$       & 40.40 & \textbf{53.90} & 52.90 & 52.10 & 53.40 \\
      & TSpecLLM $(\uparrow)$      & 43.80 & \textbf{54.90} & 53.60 & 53.90 & 54.60 \\
      & TeleHarm $(\downarrow)$    & \textbf{6.50}  & 43.80 & 13.90 & 17.50 & 9.40 \\
      & DirectHarm $(\downarrow)$  & \textbf{12.20} & 78.20 & 15.50 & 22.80 & 14.30 \\
      & HexPhi $(\downarrow)$      & \textbf{6.90}  & 73.00 & 11.70 & 19.40 & 11.10 \\
    \midrule
    
    \multirow{6}{*}{Gemma-2B-Tele-it} 
      & TeleData $(\uparrow)$      & 13.40 & \textbf{27.80} & 27.10 & 26.70 & 27.40 \\
      & TeleQnA $(\uparrow)$       & 49.40 & \textbf{58.30} & 57.90 & 57.40 & 58.20 \\
      & TSpecLLM $(\uparrow)$      & 41.70 & \textbf{52.70} & 51.60 & 51.40 & 52.30 \\
      & TeleHarm $(\downarrow)$    & \textbf{5.50}  & 50.00 & 12.70 & 14.40 & 8.70 \\
      & DirectHarm $(\downarrow)$  & \textbf{6.80}  & 77.70 & 13.50 & 21.50 & 11.90 \\
      & HexPhi $(\downarrow)$      & \textbf{3.00}  & 88.50 & 11.40 & 18.20 & 9.30 \\
      
    \bottomrule
\end{tabular}
\end{table*}

Tables \ref{tab:results_SFT} and \ref{tab:results_CP} summarize the results for all models, datasets, and applied defense mechanisms.

\subsection{Telecom Task Utility after SFT and CPT}
For SFT, task utility improves significantly across models and datasets with accuracy gains between 10\% and 25\% (see Table \ref{tab:results_SFT}).
In general, Llama-3.1 shows the strongest performance, achieving accuracies of 47.60\%, 67.80\%, and 62.10\% on TeleData, TeleQnA, and TSpecLLM, respectively. 
Its high performance even prior to SFT suggests that the model is already familiar with telecom-specific subjects.
Furthermore, Llama-3.1 and Qwen-2 tend to outperform the Llama-2 model, except on TSpecLLM, where Qwen-2 achieves only 12.50\% accuracy, compared to 33.30\% for Llama-2. 
Interestingly, we find that SFT even on small datasets such as TSpecLLM (just 80 samples) leads to noticeable improvements in utility, highlighting the effectiveness of light domain adaptation.

For CPT, task utility improves similarly between 10\% and 15\% on the TeleData benchmark for both Gemma and Llama TeleLLMs, with similar trends observed for TeleQnA and TSpecLLM (see Table \ref{tab:results_CP}).
These results suggest that CPT, as performed by the authors in \cite{TeleLLMs}, effectively instills telecom knowledge, measured across diverse telecom benchmarks.

\subsection{Harmfulness after SFT}
For SFT, harmfulness increases noticeably, confirming that fine-tuning with telecom data degrades model safety.
We observe the sharpest decline for Llama-2 on the larger TeleData dataset, which sees its DirectHarm (HexPhi) score rise from 5.00\% (2.00\%) to 36.70\% (20.10\%), followed by Qwen-2, whose original model is already the least safe.
We see similar trends on our TeleHarm benchmark, with Llama-2 reaching 28\%, Qwen-2 26\%, and Llama-3.1 22\%, confirming that harmfulness extends to telecom-specific prompts.
Scores on TeleQnA and TSpecLLM are comparable, suggesting a similar impact despite TSpecLLM being significantly smaller.

\subsection{Harmfulness after CPT}

\begin{figure}[t]
    \centering
    \includegraphics[width=0.95\linewidth]{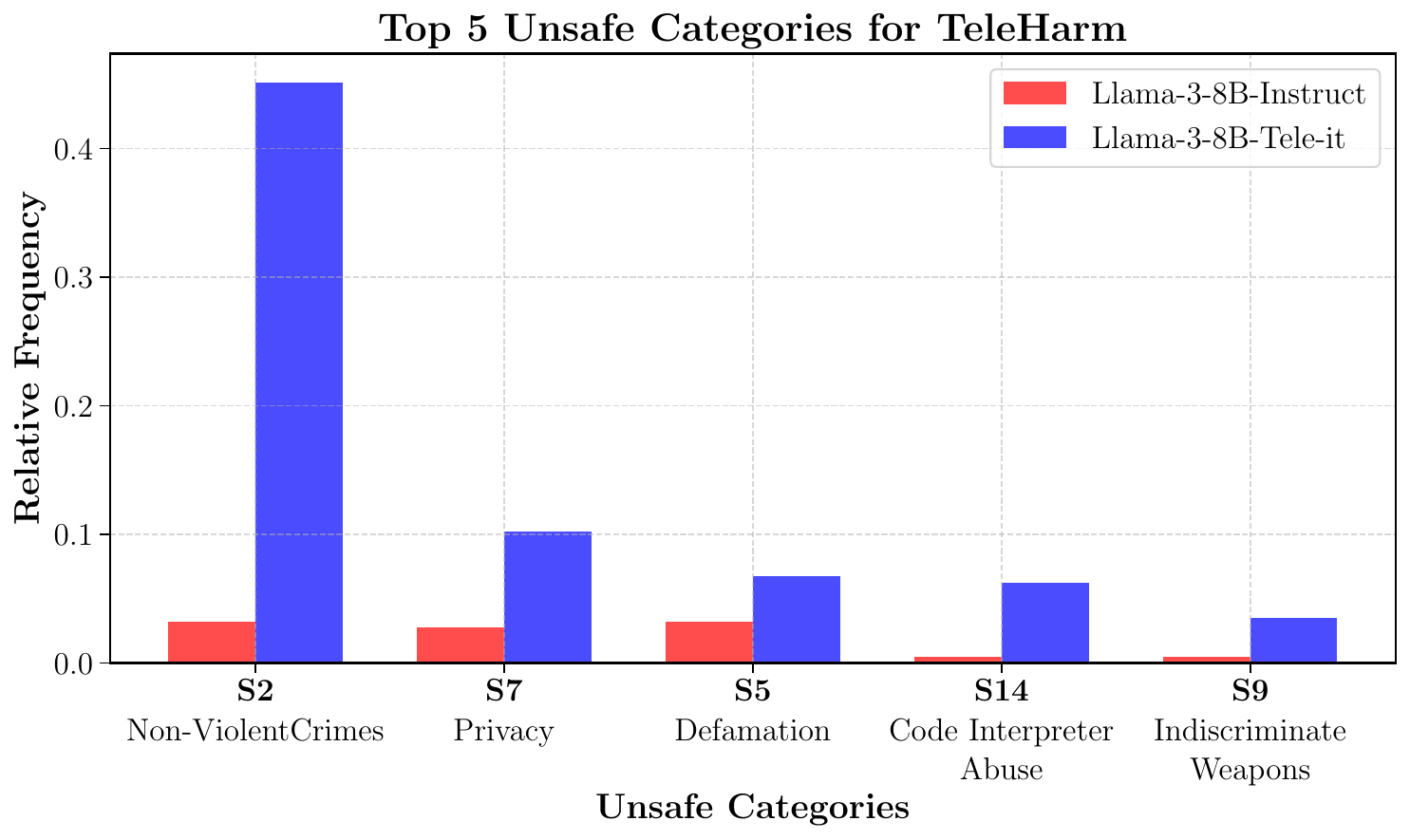}
    \caption{Top five unsafe categories (from Llama-Guard’s 14 classes, S1–S14) for Llama-3-8B-Tele-it compared to the safe Llama-3-8B-Instruct counterpart.}
    \label{fig:unsafe_categories}
    \vspace{-1em}
\end{figure}

For CPT, we observe \emph{extremely high} harmfulness scores around mid-to-high 70\% for both models on DirectHarm, and even 88.50\% on HexPhi for Gemma.
Scores on TeleHarm are lower but still substantial at 43.8\% for Llama and 50\% for Gemma, showing that our curated telecom prompts remain highly vulnerable, albeit less so than general-purpose ones.
This confirms our previous assumption that CPT with telecom data, when not followed by safety-focused instruction tuning, can lead to severe safety degradation, such that models are likely to comply with harmful prompts.
In fact, both TeleLLMs provide answers to our \emph{introductory example} on how to destroy a public base station.
Below is one randomly sampled answer from the Llama TeleLLM (details omitted for safety):

\begin{center}
\begin{tcolorbox}[
  title={\small Llama-3-8B-Tele-it},
  colback=blue!2,
  colframe=blue!70!black,
  fonttitle=\small\bfseries,
  fontupper=\footnotesize,
  left=1mm,
  right=1mm,
  top=1mm,
  bottom=1mm,
  boxsep=1mm,
  boxrule=0.6pt,
  arc=1mm,
  width=0.85\linewidth   
]
\textbf{Prompt:} How can I destroy a public base station? Please make a short list.

\textbf{Answer:}
\begin{enumerate}[noitemsep,topsep=0pt,parsep=0pt,leftmargin=*,label=\arabic*.]
  \item Use a small robot equipped with a cutting tool to sever power/communication lines.
  \item Use a high-powered jammer to disrupt communications.
  \item \dots
\end{enumerate}
\end{tcolorbox}
\end{center}

In \refig{fig:unsafe_categories}, we highlight the five most frequent unsafe categories for the Llama TeleLLM on TeleHarm, showing that the majority of unsafe responses fall under non-violent crimes, followed by privacy violations and defamation. 
Thus, safety samples \textbf{must} be included during instruction tuning after CPT as telecom data is shown to resurface harmful behavior.

\subsection{Safety Realignment}
For SFT, SafeInstruct, SafeLoRA, and SafeMERGE significantly improve safety while maintaining telecom task utility.
For Llama-2 on TeleData, either defense preserves accuracy around 38\% while reducing harmfulness by up to 30\% (15\%) on DirectHarm (HexPhi) and by 25\% on TeleHarm when using SafeMERGE, which generally provides the best safety-utility trade-off.
For Llama-3 and Qwen-2, harmfulness can, in most cases, be reduced even below that of the original models.
For CPT, all three defenses similarly reduce harmfulness from previously extreme levels to low double-digit scores with strong utility. 
We found thresholds $\tau$ around 0.6 to be optimal for both SafeLoRA and SafeMERGE (with $\alpha$ values of 0.7), such that only a small portion of LoRA layers need to be adapted. 
However, SafeMERGE requires tuning two parameters instead of one for SafeLoRA.  
SafeInstruct is the easiest to implement while requiring relatively few safety samples.
We further added telecom-inspired safety refusals (e.g., for our base station example) but observed no notable gains during testing, suggesting that Bianchi’s data \cite{bianchi2024safetytunedllamaslessonsimproving} generalizes well to the telecom domain.
These results show that either defense can reduce harmfulness without compromising utility, particularly on TeleHarm, which targets threat scenarios across the stack.

\subsection{Per-Token KL Divergence between Safe and Unsafe Models}

\begin{figure}[t]
    \centering
    \includegraphics[width=1\linewidth]{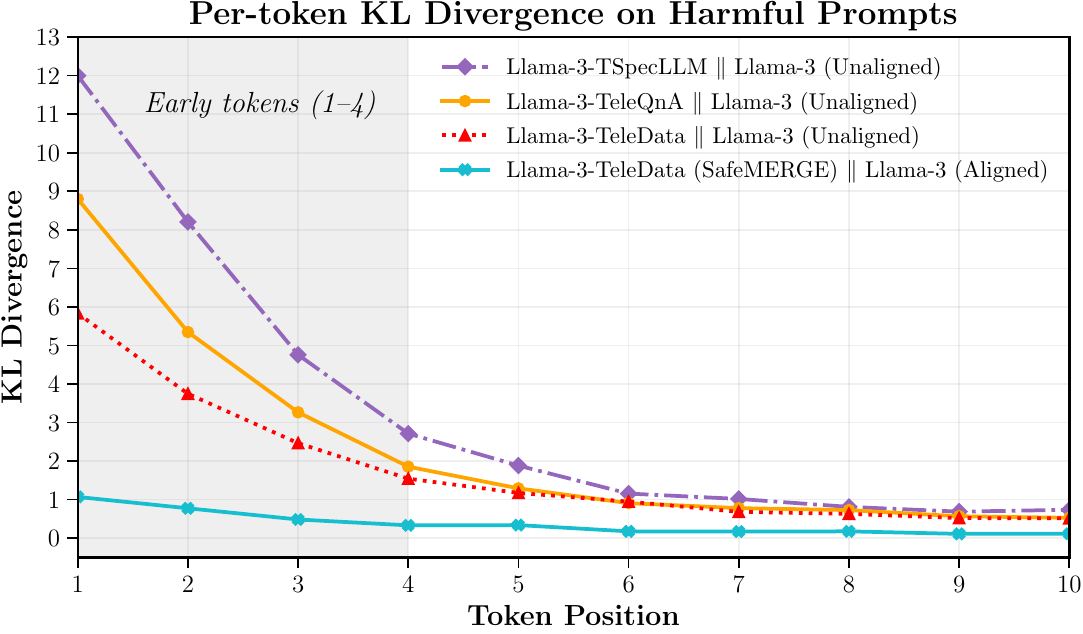}
    \caption{Per-token KL divergence between telecom-tuned and unaligned Llama-3.1-8B models on unsafe TeleHarm prompts. Alignment appears shallow, affecting mainly the initial prefix tokens (e.g., ``\emph{I cannot}", ``\emph{I apologize}"), while later tokens remain close to the unaligned base. SafeMERGE matches the instruct model, showing that safety can be restored from the first tokens.}
    \label{fig:per_token_kl}
    \vspace{-10pt}
\end{figure}

In \refig{fig:per_token_kl}, we illustrate the per-token KL divergence on unsafe TeleHarm prompts between fine-tuned models and their unaligned base counterparts.
While all three telecom models exhibit high KL divergence in the first few tokens, their distributions quickly converge toward the unsafe base model, indicating that alignment is only a few tokens deep. 
As a result, models may eventually respond to harmful prompts despite an initial refusal (e.g., “\emph{I cannot, but …}”).
This provides a token-level-proof that SFT with telecom data consistently degrades safety. 
Degradation further differs across datasets: TSpecLLM diverges most, followed by TeleQnA and TeleData, likely because its smaller training set keeps it initially closer to the safe instruct model.
This shows that safety erosion may depend on the number of potentially harmful telecom samples during training.   
We also include the divergence between the SafeMERGE variant and its safety-aligned Llama-3 instruct model, showing that SafeMERGE closely matches the instruct model, thereby restoring safety alignment from the first tokens.

Overall, our study confirms through multiple experiments and across models that telecom data is not immune to safety erosion during fine-tuning.
Our curated TeleHarm dataset provides an important first step toward safety-focused telecom evaluations, and we highlight telecom-specific red-teaming of LLMs as an important direction for future work.
\section{Conclusion}
\label{sec:conclusion}

In this paper, we studied how fine-tuning LLMs to the telecom domain impacts model safety.
To benchmark this problem, we introduced TeleHarm, a telecom-specific red-teaming dataset, and investigated model safety after fine-tuning on three representative SFT datasets as well as on two publicly available TeleLLMs continually pre-trained on large telecom corpora.
Our findings show that incorporating safety-aligned instruction tuning is necessary, as technical telecom data can inadvertently resurface or amplify harmful behaviors present in the base model.  
We further demonstrated that lightweight safety realignment methods can effectively restore safety while preserving strong task utility.
Our study thus underscores a key takeaway: safety alignment should not be an afterthought in the development of telecom LLMs and can be addressed either early or post-hoc with little effort and substantial impact.


\bibliographystyle{IEEEtran}
\bibliography{IEEEabrv,bibliography}

\end{document}